# Realistic 3D computer model of the gerbil middle ear, featuring accurate morphology of bone and soft tissue structures

**Jan A.N. Buytaert**[1,°]**, W.H.M. Salih**[1]**, M. Dierick**[2]**, P. Jacobs**[2] **and Joris J.J. Dirckx**[1]

[1] Laboratory of BioMedical Physics – University of Antwerp, Groenenborgerlaan 171, B-2020 Antwerp, Belgium
[2] Centre for X-ray Tomography - Ghent University Proeftuinstraat 86, B-9000 Gent, Belgium

Running title: Accurate 3D gerbil ME model

° Corresponding author:
  email jan.buytaert@ua.ac.be;
  telephone 0032 3 265 3553;
  fax 0032 3 265 3318.

47    ABSTRACT

48

49    In order to improve realism in middle ear (ME) finite element modeling (FEM),

50    comprehensive and precise morphological data are needed. To date, micro-scale X-ray

51    computed tomography (μCT) recordings have been used as geometric input data for FEM

52    models of the ME ossicles. Previously, attempts were made to obtain this data on ME soft

53    tissue structures as well. However, due to low X-ray absorption of soft tissue, quality of

54    these images is limited. Another popular approach is using histological sections as data for

55    3D models, delivering high in-plane resolution for the sections, but the technique is

56    destructive in nature and registration of the sections is difficult.

57    We combine data from high-resolution μCT recordings with data from high-resolution

58    orthogonal-plane fluorescence optical-sectioning microscopy (OPFOS), both obtained on the

59    same gerbil specimen. State-of-the-art μCT delivers high-resolution data on the three-

60    dimensional shape of ossicles and other ME bony structures, while the OPFOS setup

61    generates data of unprecedented quality both on bone and soft tissue ME structures.

62    Each of these techniques is tomographic and non-destructive, and delivers sets of

63    automatically aligned virtual sections. The datasets coming from different techniques need

64    to be registered with respect to each other. By combining both datasets, we obtain a

65    complete high-resolution morphological model of all functional components in the gerbil

66    ME. The resulting three-dimensional model can be readily imported in FEM software and is

67    made freely available to the research community.

68    In this paper, we discuss the methods used, present the resulting merged model and discuss

69    morphological properties of the soft tissue structures, such as muscles and ligaments.

70

71

72

77

78

79

80    INTRODUCTION

81

82    The middle ear (ME) forms a small three-dimensional (3D) biomechanical system. It mainly

83    consists of the tympanic membrane (TM), three ossicles – malleus, incus and stapes – and

84    their supporting ligaments and muscles. The remarkable performance of ME mechanics is

85    too complex to be understood intuitively. For better understanding, ME modeling was

86    introduced. Finite-element (computer) modeling (FEM) has become an established

87    numerical technique to simulate ME mechanics. In ME research, the technique was first

88    introduced by Funnell and Laszlo, 1978. As one of its inputs, FEM requires 3D morphological

89    computer models of the ME components. These mesh models consist of a finite number of

90    elements, e.g. tetrahedra or hexahedra.

91

92    Current morphological models are either incomplete, low resolution and/or contain

93    rudimentary shapes to represent (some) ME components. Pioneering work in this field used

94    manually drawn geometrical shapes in the computer to represent the ME malleus, incus and

95    stapes (Wada et al., 1992; Ladak and Funnell, 1996; Blayney et al., 1997; Eiber et al., 2000;

96    Prendergast et al., 2000; Koike et al., 2002). Some authors used low or modest resolution

97    shapes measured with medical X-ray computed tomography (CT) (Rodt et al., 2002; Lee et

98    al., 2006) or with tabletop micro-CT (μCT) devices (Decraemer et al., 2002; 2003; Elkhouri et

99    al., 2006; Puria and Steele, 2010; Lee et al., 2010). Other authors used histological sectioning

100   (Funnell et al., 1992; Sun et al., 2002) or magnetic resonance microscopy (MRM, NMR, MRI)

101   (Funnell et al., 2005; Elkhouri et al., 2006), but again with modest resolutions. In many

102   models, the suspensory ligaments and muscle tendons are either omitted (Wada et al. 1992;

103   Ladak & Funnell 1996; Blayney et al. 1997; Lord et al. 1999; Rodt et al. 2002) or manually

104   incorporated as simple geometrical objects such as blocks, cylinders or cones  (Prendergast

105   et al. 2000; Beer et al. 2000; Koike et al. 2002; Sun et al. 2002; Lee et al. 2006). To the

106   authors' knowledge, only models by Wang et al., 2006; Gan et al., 2007; Cheng and Gan,

107   2008 (using histological sectioning) and by Mikhael et al., 2004; Sim and Puria, 2008; Ruf et

108   al., 2009 (using X-ray techniques) contain actually measured shapes of soft tissue structures,

109   but in low resolution.

110

111   To improve realism in FEM calculations, ME geometry models need to incorporate all and
112   accurate shapes of the ossicles and suspensory soft tissue structures (Decraemer et al.,
113   2003). As the computer calculating capacity has grown to a point where it can manage large
114   amounts of data, and as the scientific measurement apparatus is now capable of high-
115   resolution imaging on all kinds of tissue types, the time has come to incorporate realistic and
116   complete morphological 3D ME models in FEM. We point out that it might not be necessary,
117   even not numerically feasible, to perform FEM with all structures described in the highest
118   detail. On the other hand, it is difficult to decide beforehand how precise the morphologic
119   model needs to be. Therefore, we think it is important to first have a high-resolution
120   morphologic model available, which can then be simplified to the modeler's judgment.

121

122   In the current paper we provide these high-quality models by combining data originating
123   from two different tomographic techniques: State-of-the-art μCT tomography allows to
124   obtain precise data on bony structures, but due to the low X-ray absorption of soft tissue, CT
125   generates poor quality images of soft tissue (Lemmerling et al., 1997). Therefore we
126   combine these data with measurements from another and relatively new technique:
127   Orthogonal-plane fluorescence optical-sectioning (OPFOS) microscopy or tomography. This
128   method images both bone and soft tissue at the same time and in high-resolution.

129

130   As gerbil is one of the standard laboratory animal models in fundamental hearing research,
131   we chose this species for our first model.

132

133

134   MATERIALS and METHODS

135

136        Dissection

137        All animal manipulations in this work were performed in accordance with Belgian
138   legislation and the directives set by the Ethical committee on Animal Experimentation of our
139   institution (University of Antwerp, Belgium). Three adult Mongolian gerbils (*Meriones*
140   *unguiculatus*), aged between three and six months, were used. They were housed in cages
141   with food and water ad libitum in our animal facility.

142

143   The animals were euthanized using carbon dioxide, followed by a cardiac perfusion with
144   physiological fluid to rinse out all the blood from the gerbil head blood vessels. This step is
145   necessary to allow for OPFOS tomography (as we will explain below). The gerbils are then
146   decapitated and the right temporal bones were isolated. The specimens were reduced in size
147   until only the bulla was left containing the middle and inner ear, cf. Figure 1. During the
148   harvesting of these bullas, continuous moistening with mist from an ultrasonic humidifier
149   (Bionaire BT-204) was applied to avoid dehydration.

150

151

152       Cross sectional imaging of bone with X-ray tomography

153       The first stage of 3D tomographic recording of the ME was achieved using micro-scale
154   X-ray computed tomography. The dissected bullae were enclosed in separate Eppendorf
155   vials, together with a calibration object and a few droplets of physiological fluid at the
156   bottom. In this way a 100% saturated humid environment was created to avoid dehydration
157   artifacts. Another droplet of fluid was placed in the ear canal – which could help – to
158   distinguish the outline border of the TM shape with the air-filled ME cavity. Water and air
159   have a slightly different X-ray absorption coefficient, so a layer of water on the extremely
160   thin TM can help to reveal its medial shape outline. In previous work, we measured the
161   shape of the eardrum before and after putting fluid on the membrane: even with a 10 mm
162   water column in the ear canal, no measurable deformation was found with moiré
163   profilometry of 15 μm resolution (Buytaert et al., 2009). As the droplet of water is less than 3
164   mm high (inducing a pressure load of 30 Pa) the TM deformation is well below the μCT
165   measurement resolution. The Eppendorf vials (made from polypropylene) are almost X-ray
166   transparent. Especially bone absorbs X-rays well, thus creating a high contrast in
167   transmission recordings. The small calibration objects were custom-made from polyvinyl
168   chloride (PVC) in our mechanical workshop and possess about the same X-ray absorption
169   properties as thin bone (Gea et al., 2005). They served as an independent calibration to
170   verify the μCT device specifications.

171

172   The vials containing gerbil specimens were scanned at the UGCT scanning facility at Ghent
173   University (www.ugct.ugent.be) using a custom-built μCT scanner of medium energy (up to
174   160 keV). The scanner has a directional X-ray tube with a feature recognition capability up to

175    2 μm (Masschaele et al., 2007). The scans were performed at a tube voltage of 120 kV
176    (photon energy levels ranging from 0 to 120 keV) and a current of 58 μA. A custom-made vial
177    holder was mounted on a computer-controlled rotation table (MICOS, UPR160F-AIR). For
178    each specimen a series of 1000 shadow projections of 1496x1880 pixels was recorded
179    covering an object rotation of 360 degrees (or one recording every 0.36 degrees).
180    Reconstruction of the tomographic data volume to serial sections was achieved using the
181    back-projection algorithms of the Octopus software package (Dierick et al., 2004), resulting
182    in 1780 reconstructed cross sections of 1496x1496 pixels. From these calculated cross
183    sections with an isometric pixel size of 8.5 μm, accurate 3D models of the three ossicles and
184    other bony structures were generated. All three datasets cover a volume of 15.1 x 12.7 x
185    12.7 mm (1780x1496x1496 times 8.5 μm).

186
187
188            Cross sectional imaging of soft tissue with optical tomography

189            Due to the low X-ray absorption of soft tissue, another tomographic technique was
190    needed: orthogonal-plane fluorescence optical-sectioning microscopy or OPFOS (Voie et al.,
191    1993). OPFOS was initially developed to image the inner ear cochlea, but it has also been
192    used in ME studies (Voie, 2002; Buytaert and Dirckx, 2007; 2009). In the OPFOS method,
193    parallel optical sections through a macroscopic biomedical specimen are created by means
194    of a thin sheet of laser light, and the fluorescence originating from within the cross section of
195    the light sheet with the tissue is recorded in the direction perpendicular to the plane of the
196    laser light. The light emitted by the specimen originate from auto-fluorescence or from
197    staining the specimen with a fluorescent dye. OPFOS images both bone and soft tissue at the
198    same time and in real-time, as no (back-projection) calculations are required. It allows
199    region-of-interest (ROI) imaging and has both a high sectioning and a high in-plane
200    resolution. Hence, perfectly and automatically aligned images of virtual cross sections can be
201    obtained. OPFOS scanning was performed at the Laboratory of BioMedical Physics at the
202    University of Antwerp (www.ua.ac.be/bimef) with a custom-built setup using bi-directional
203    light-sheet illumination (Buytaert and Dirckx, 2009; Buytaert, 2010).

204
205    For OPFOS imaging, an elaborate specimen preparation is needed (Voie, 2002; Buytaert and
206    Dirckx, 2009), as the technique requires the specimens to be perfectly transparent. Before

207    µCT scanning, all blood was removed from the blood vessels, as coagulated blood cannot be

208    made transparent afterwards. After µCT recording, a 10% neutral buffered Formalin bath

209    was applied. Next, all calcium was removed using 10% EDTA in water solution combined with

210    microwaves. Because of this decalcification, the OPFOS method has to be performed second

211    after µCT X-ray scanning. Then, the specimens were dehydrated using a slowly graded

212    ethanol series, up till 100%. Next, all tissue was refractive index matched using a slowly

213    graded Spalteholz fluid series, again up till 100%. As a result, the specimens become entirely

214    transparent when submerged in pure Splateholz fluid. Finally, to obtain stronger

215    fluorescence, the specimens are stained with Rhodamine B.

216

217    Both soft tissue and bone were made transparent and fluorescent; hence, both tissue types

218    are visualized with the technique. We focused on region-of-interest (ROI) OPFOS imaging of

219    ME ligaments, tendons and muscles, while images of the (often larger) bony structures are

220    more easily obtained from µCT. Comparison of high-resolution µCT and OPFOS data allows

221    us to distinguish bone from soft tissue in the OPFOS data. Merging of the two datasets

222    generates the complete ME model with all of its functional components accounted for.

223

224    The shape of the TM was obtained from the µCT data. The OPFOS technique is able to

225    visualize this extremely thin tissue when performing ROI imaging on a small part of the

226    membrane, cf. Figure 2. However, to image the membrane full-field with OPFOS, one needs

227    to zoom out and the resolution needed to adequately visualize this thin membrane is lost.

228    Furthermore, the eardrum is prone to preparation artifacts: Because the gerbil specimens

229    went through an extensive procedure of tissue fixation, decalcification, dehydration and

230    Spalteholz treatment, the extremely thin TM can get deformed. Therefore, the data on

231    eardrum shape are obtained from the CT images, recorded before any specimen processing

232    was applied. X-rays are normally not suited to image soft tissue, especially if it is very thin,

233    like the eardrum. We tried to counter this problem by applying a droplet of physiological

234    fluid through the ear canal on top of the membrane. The medial border of the droplet and

235    eardrum then become more easily distinguishable from air in the ME cavity. In this way, the

236    membrane outline will be obtained without deformation and with adequate resolution.

237

238    Apart from the specimen preparation, the OPFOS method has another disadvantage as it

239   suffers from stripe artifacts. Opaque regions or areas of less transparency locally reduce the
240   intensity of the laser light sectioning sheet, causing shadow lines or stripes in the rest of the
241   image. This is partially countered by simultaneous dual light sheet illumination in our setup
242   (Buytaert and Dirckx, 2009).

243

244   Measuring and analyzing the OPFOS data is very time-consuming; therefore, only one gerbil
245   ear has been processed. On the other hand, the μCT data of all three gerbils was analyzed.

246

247

248        Visual observations

249        We performed visual observations of the orientation, location, shape and suspension
250   of the ossicular chain inside opened ME bullae with an operating microscope (Zeiss, OPMI
251   Sensera S7). When 3D computer data, models and results were obtained from μCT or OPFOS
252   with striking features, they were compared to qualitative observations of the real geometry
253   in opened bullae with the operating microscope to verify their interpretation. These
254   experiences gave us the necessary expertise to confirm the 3D model results and conclusions
255   of the present paper. For instance, after a targeted dissection we could visually confirm that
256   the posterior incudal ligament in gerbil indeed exists as one whole band instead of two
257   separate structures, as we found in our OPFOS data and model.

258

259

260        3D segmentation and reconstruction

261        After obtaining several series of object cross sections – one μCT set originating from
262   back projection calculations, and several ROI datasets from direct OPFOS recordings – we
263   identified and segmented the relevant structures in all images. The goal of segmentation is
264   to locate objects boundaries, which in turn allows software to build 3D surface meshes by
265   triangulation.

266

267   In our case, segmentation was done manually for thousands of sections using the
268   commercial image segmentation and three-dimensional surface mesh generating software
269   package Amira 5.3 (Visage Imaging). Manual segmentation might seem primitive and time-
270   consuming; but using our morphological expertise, manual segmentation delivers better
271   results than purely automated segmentation based on thresholding of gray scale values. The

272    Amira software package uses the marching cubes algorithm for triangulation. It takes eight
273    neighboring voxel locations at a time (forming an imaginary cube), after which the
274    polygon(s) needed to represent the part of the isosurface that passes through this cube are
275    determined. The individual polygons are finally fused into the intended surface. This leads to
276    subvoxel triangulation that easily manages sharp angles. When smoothing or simplification
277    (reduction of the number of triangles) is used, the program takes the 'steepness' of the
278    surface into account: Flat surface parts are more reduced than curved parts.
279
280    As final result, we end up with triangulated surface meshes for the μCT and OPFOS datasets.
281    These can be further developed into finite-element volume meshes using Amira or other
282    packages. On the website of the Laboratory of BioMedical Physics group we suggest some
283    powerful and open-source volume generating software, e.g. PreView.
284
285
286        Merging of CT and OPFOS models

287        All cross sections in a μCT dataset, and therefore all models of ME components
288    originating from it, are inherently perfectly aligned within the data stack. The OPFOS
289    datasets were focused on the soft tissue by separate ROI recordings. However, parts of the
290    bone are included in the OPFOS ROI recordings as well. The cross sections within each ROI
291    OPFOS data stack are also perfectly aligned, but the resulting mesh models per stack are
292    unrelated to the other OPFOS datasets (because of different ROI zooming and/or other
293    slicing orientation) and unrelated to the CT dataset and models.
294
295    To merge the OPFOS data with the μCT data, the μCT dataset was used as a reference. We
296    did not merge the 2D image cross sections, but the 3D mesh models: All partial bone models
297    from ROI OPFOS were three-dimensionally aligned to corresponding parts of the μCT models
298    using an iterative spatial transformation least-squares minimization process of the Amira
299    software package. This process uses the Iterative Closest Point (ICP) algorithm to minimize
300    the difference between two point clouds (e.g. all surface nodes of respectively an OPFOS and
301    a μCT mesh model). ICP iteratively revises the spatial transformation (6 degrees of freedom
302    for translation and rotation) needed to minimize the Euclidean distance between the points
303    of two datasets. This concept is referred to as the Procrustes superimposition method: The

304   root-mean-square (RMS) of the distances between corresponding points of the two surfaces
305   are evaluated. Corresponding point pairs are created by finding the closest point of the
306   reference (μCT) surface mesh for *each* point of the other (OPFOS) surface mesh. When the
307   two surfaces are identical and perfectly superimposed, the RMS of all corresponding point
308   distances will be zero. In the case of the OPFOS versus the μCT stapes model for instance, we
309   obtained a root-mean-square difference of 17 μm (or two μCT voxels). After obtaining such a
310   good match between the OPFOS and μCT bone model, we applied the same spatial
311   transformation to the OPFOS soft tissue mesh(es) from that OPFOS dataset. In this way, all
312   OPFOS datasets were combined with μCT data into one model.

313
314

315   RESULTS

316

317       Computed tomography

318       Three gerbil ears were recorded with μCT, delivering three isometric data stacks of
319   reconstructed cross sections (pixel size 8.5x8.5 μm, separated 8.5 μm). To illustrate the
320   image quality, we present one μCT cross section in Figure 3. Full movies of all cross sections
321   are available on our website and the entire dataset is available upon request. Notice how
322   distinguishable the ossicle boundaries, the incudomallear and incudostapedial joint cleft and
323   the annular ligament cleft are in the figure. This high contrast and resolution facilitates the
324   segmentation process considerably.

325

326   Our main attention went to the ME; but separate 3D surface meshes were also created of
327   the fluid-filled bony labyrinth of the inner ear (cochlea scalae and modiolus, and vestibular
328   apparatus), cf. Figure 1 and Figure 4. The ME bulla air cavities of all gerbils are modeled as
329   well. They give an indication of the enclosed air volume in the ME, cf. Table 1. These
330   segmented volumes include the volume of the ossicles, ligaments and muscles. Finally, a
331   separate rudimentary mesh of all bone using a fixed segmentation threshold was made.
332   Using transparent rendering for this large model, one can virtually look inside the bulla and
333   observe the ossicles and inner ear inside, cf. Figure 1. We listed volume, dimensions and
334   several other properties of the ossicles, the TM and the ME bulla cavity in Table 1 to Table *3*.
335   These and other quantitative data is readily and accurately available from our models.

336   The mass of malleus, incus and stapes are respectively 1.145 mg, 0.633 mg and 0.116 mg as

337   reported by Nummela, 1995. Adopting these representative values for our specimen in

338   combination with the volumes given in Table 1, we get an average ossicle bone density of

339   $1.37 \times 10^3$ kg/m³ for the stapes and $1.74 \times 10^3$ kg/m³ for incus and malleus.

340

341   Note that the outline of the TM was surprisingly but successfully visualized using µCT. The

342   resolution was just high enough to show the shape outline of the extremely thin membrane.

343   Thickness information could not be obtained. Using a fluid droplet in the ear canal to aid in

344   distinguishing the medial border of the eardrum partially failed, as can be seen in Figure 3:

345   Fluid is not covering the entire membrane surface in the ear canal because of an air bubble.

346

347   Finally, we could observe channels (blood vessels) inside the ossicles, occurring especially in

348   the incus and malleus bone, cf. Figure 5. The ossicular surface shapes are almost identical

349   between all three animals, and the same is true for the size, volume and branching layout of

350   the major channels inside them.

351

352

353   OPFOS tomography

354   We will now discuss all identified (soft) tissue structures of the ME of gerbil 2,

355   measured with OPFOS.

356

357   *Posterior incudal ligament*

358   Using µCT, the posterior incudal ligament cannot be found, cf. Figure 6A-B, while using

359   OPFOS it is clearly visible, cf. Figure 6C-E. This comparison between the two tomographic

360   imaging techniques clearly demonstrates the usefulness of combining the two methods.

361

362   After segmentation and 3D representation, cf. Figure 6F-G, one can see that the ligament is

363   built as one whole part and forms one sickle-shaped band of fibrous tissue. Its tiny volume

364   amounts to 0.013mm³. The sickle has its smallest thickness (orthogonally to the image plane

365   of Figure 6F) of 42 µm near the incus short process and broadens to 190 µm towards the

366   bulla edge. The contact area at the middle-ear cavity wall is also a bit larger than the contact

367   area on the incus crus.

368

369                      *Anterior mallear ligament*

370    The anterior process of the malleus has the shape of a (partially opened) hand-held Japanese

371    folding fan, reaching towards the anterior bulla wall, cf. Figure 7 and Figure 8. The

372    connective soft tissue of the anterior mallear ligament, which should connect the process to

373    the bulla, is undistinguishable from bone, both in the OPFOS as in the CT recordings. This

374    ligament is probably more ossified or cartilaginous than in some other species and no

375    separate soft tissue model could be made.

376

377              *Superior mallear and incudal ligament and lateral mallear ligament*

378    According to real-time OPFOS observations, no superior mallear and incudal ligament are

379    present in the gerbil ME, which is confirmed by visual observations with the operating

380    microscope. In addition, no lateral mallear ligament could be discerned with either method.

381

382                      *Tensor tympani muscle and tendon*

383    Figure 2A shows a high-resolution OPFOS section image through the tensor tympani muscle

384    and tendon, the TM, the malleus' manubrium and the bulla. This image demonstrates

385    OPFOS' capability to image bone and soft tissue in high-resolution.

386

387    After the segmentation and triangulation process, the volume of the tensor tympani muscle

388    and tendon can be calculated from the obtained 3D model, and was found to be 0.486 mm³.

389    The distance between the two most distant points on the combined structure is 3.25 mm.

390    The diameter of the muscle tendon varies between 50-80 μm.

391

392    The cross-sectional area of a muscle (rather than volume or length) determines the amount

393    of force it can generate. A first rough estimate of the order of magnitude of the maximum

394    generated force of the muscle can be derived as follows. By dividing the muscle belly volume

395    by an average muscle fiber length of 400 μm (estimated from the OPFOS images), we end up

396    with a cross sectional area of $1.2 \times 10^{-2}$ cm². A common conversion factor from this area to

397    the maximal isometric contraction force is given by 25 N/cm² for skeletal muscle (Nigg and

398    Herzog, 1999), giving a maximally generated force of this muscle of 0.3 N. An interesting

399    comparison of the effect of this force on the malleus can be made by translating this number

400 into a corresponding static pressure working on the TM from the ear canal side. Dividing the
401 force of 0.3 N by the (projected) area of the pars tensa of the TM of gerbil 2 (13.64 mm², cf.
402 Table 2), we obtain a (maximum) static pressure of 22 kPa. The magnitude of this pressure
403 falls in the range of static pressures associated with scuba diving or taking an airplane.

404

405 The final merged 3D model shows that the tensor tympani muscle belly is larger than
406 expected from visual observations. Its main part is hidden as it is situated in a gap between
407 the spiraled cochlear dome and the bulla wall.

408

409    *Stapedial artery*

410 A typical anatomical feature of the gerbil ME is the stapedial artery running through a bony
411 channel on the surface of the first cochlear turn and passing through the stapes crura in the
412 ME air cavity. Using OPFOS, it was possible to image this relatively large stapedial artery, cf.
413 Figure 9. We could even distinguish and separately model the stapedial artery soft tissue
414 wall (the actual blood vessel) and its fluid-filled lumen.

415

416 The diameter of the blood vessel was the smallest in between the crura and amounted to
417 355 µm with (i.e. outer diameter) and 275 µm without (i.e. inner diameter) the blood vessel
418 soft tissue wall. The wall had a thickness of about 40 µm.

419

420    *Stapedius muscle and tendon*

421 After segmentation of the stapedius muscle and tendon, we end up with the mesh shown in
422 Figure 9. The tiny volume enclosed in this (tendon & muscle) mesh amounts to 0.085 mm³
423 and the two most distant points on the combined structure are 1.81 mm apart. The diameter
424 of the tendon varies between 40-55 µm. If we again divide the volume by an estimated
425 average muscle fiber length of 350 µm, we get a cross sectional area of $2.4 \times 10^{-3}$ cm².
426 Multiplying this value by 25 N/cm² gives an estimation of 0.06 N for the maximum force the
427 muscle can produce.

428

429 The merged 3D model shows that the stapedius muscle body is attached to the lateral
430 (horizontal) semi-circular canal, cf. Figure 9. In the figure, a gap is seen between the semi-
431 circular canal and the muscle because only the fluid-filled cavity of the canal is shown. When

432   showing bone as well, one sees the muscle clasps firmly around the lateral semi-circular
433   canal wall.

434

435           *Joint clefts*

436   As can be seen in Figure 2B, the incudomallear and incudostapedial joints can be easily
437   distinguished on high-resolution OPFOS cross sections and appear to form a tight
438   connection. μCT data also show both clefts, from which we made three-dimensional meshes.

439

440   The incudomallear joint connects the incus and malleus and has the shape of a twisted
441   saddle. The gap or cleft between the ossicles could contain synovial fluid as it is considered a
442   synovial joint; however, this is not confirmed from our OPFOS measurements nor μCT data
443   in gerbil. No fluid or open space is detected in the joint cleft, and the joint seems quite rigid.
444   This rigidness was already reported for other species by Guinan and Peake, 1967; Gundersen
445   and Høgmoen, 1976. The thickness of the joint varies from nearly zero to 51 μm. The gap or
446   joint tissue is thinner at the lateral side.

447

448   The incudostapedial joint connects the incus lenticular process with the head of the stapes.
449   Our model of this synovial joint shows an oval disk with an approximately even thickness of
450   25.5 μm. Again, the joint cleft seems to possess no synovial fluid and forms a rigid
451   connection, which has also been reported in cat (Funnell et al., 2005).

452

453   OPFOS also visualized the stapedial annular ligament cleft in which the annular stapedial
454   ligament is situated, forming a syndesmosis joint. A syndesmosis is a slightly movable
455   articulation where bony surfaces are tightly united by a fibrous tissue ligament (Laurent,
456   1998). The high resolution of the OPFOS data allows to make a 3D mesh of this thin
457   structure, cf. Figure 10. The thickness of the ligament varies between 8 to 18 μm, confirmed
458   by the gap seen in the μCT cleft model which is about 12-18 μm.

459

460           *Chorda Tympani*

461   The chorda tympani nerve branches from the facial nerve and runs through the ME air
462   cavity. In gerbil, the nerve jumps from a sort of support beam at the superior bulla wall to
463   the malleus where it is tightly connected to the malleus neck in the vicinity of which the

464  tensor tympani muscle connects as well, cf. Figure 7. It hangs in the ME air space passing the
465  incudal long process laterally and the manubrium medially. It rounds the malleus neck from
466  the posterior to the anterior side, passing the tensor tympani tendon inferiorly. At the
467  anterior side, it lies on the anterior process sheet until it disappears in a fissure of the bulla
468  wall again. It was unexpected that the chorda tympani could be visualized so well in OPFOS
469  cross sections, cf. Figure 8, because myelin nerve sheets can in principle not be made
470  transparent by the Spalteholz process. Apparently, because the nerve is thin enough, the
471  blurring effect of the less transparent chorda tympani was negligible.

472

473

474      Merging of CT and OPFOS models

475      As described before, we obtained a series of cross sectional images from μCT with
476  bone only, cf. Figure 3, and from OPFOS with bony and soft tissue structures, cf. Figure 2,
477  Figure 6 and Figure 8. With OPFOS, we performed ROI recording of all soft tissue structures,
478  so only incomplete parts of the ossicles were measured. However, using these partial models
479  of ossicles and/or bulla bone that were recorded together with the soft tissue, we could
480  align these bony structures (and thus the soft tissue structures as well) to the μCT bone
481  models, cf. Figure 11, using the Procrustes superimposition method.

482

483  The merging and alignment of bony structures revealed that some shrinking of the gerbil 2
484  specimen had occurred despite of our careful efforts during preparation. Using the warping
485  procedure in Amira (similar to the Procrustes superimposition method, only allowing for
486  scaling in every dimension as well), we found a shrinking factor of 8.4% in all three
487  dimensions. After applying the spatial transformation and up-scaling, the OPFOS soft tissue
488  meshes fit rather well in between the CT bone mesh models. For instance, corresponding
489  bony parts of the malleus from OPFOS using a scale-factor of 8.4% were aligned with the
490  malleus from μCT. After applying the same scaling and spatial transformation to the tensor
491  tympani, its tendon attaches to the malleus, cf. Figure 11, and at the other side its muscle
492  body inserts nicely in a bony cavity of the bulla of the inverse shape, cf. Figure 12. This and
493  similar facts give us confidence in the merging of the data.

494

495

496    DISCUSSION

497

498         Imaging method

499         Several methods exist to measure and image the ME for the creation of FEM models.

500    µCT in itself is mainly suited to image the bony structures. µCT using contrast agents is a

501    valuable alternative to our combined approach (Metscher, 2009). However, it is difficult to

502    discriminate between bone and soft tissue, so it would be necessary to do µCT scans before

503    and after staining, and merge the data as we now did with OPFOS. OPFOS offers a resolution

504    down to 2 micrometer, which is seldom achieved in µCT. For this reason, we preferred

505    OPFOS to obtain the soft tissue data. Multiple energy CT techniques have also proven to be a

506    valuable method for discriminating between soft tissue and bone in CT images (Johnson et

507    al., 2007; Granton et al., 2008). For large macroscopic structures the technique is indeed

508    feasible, however, it becomes more difficult in the case of microscopic samples. The position

509    of the micro-focus spot changes in an X-ray tube when its energy or source is altered. As a

510    result, the datasets are slightly shifted in a complicated way, and tissue discrimination can

511    no longer be done by simple subtraction or division. Gradually, these technical issues are

512    being solved, so in the future dual-energy CT may be used to measure and discriminate soft

513    tissue and bone.

514

515    The most used alternative to our method is conventional histological sectioning, which is

516    unsurpassed in resolution and produces data on the bone and soft tissue simultaneously.

517    Both the histological method as our combined method need a similar specimen preparation

518    that can induce shrinking (Lane and Ráliš, 1983; Henson et al., 1994). Our method is

519    considered non-destructive (as multiple measurements can be done on the sample) while

520    histology can only measure the sample once and in one slicing orientation because of the

521    need for physical cutting of the specimen. Furthermore, these 2D slices are often deformed

522    during slicing, requiring difficult image processing and registration of all slices before

523    generating a 3D model. µCT and OPFOS each deliver perfectly and automatically aligned

524    cross sections that require no post-processing. Instead of registering every 2D slice, our

525    method only needs to register complete 3D meshes of all submodels to one another. OPFOS

526    further allows real-time virtual sectioning and imaging.

527

528

### Human versus gerbil

530    When using animal models, it is important to be aware of the differences with human
531    ME morphology. Figure 13 shows a schematic representation of all human ME components.
532    In addition to the data prepared in this paper, we confirmed our findings in other gerbil ears
533    during other studies using OPFOS and visual inspection with the operation microscope.

534

535    We found that in gerbil no superior incudal, no superior mallear and no lateral mallear
536    ligament are present, contrary to the case in humans. The presence and/or function of
537    superior attachments to malleus and incus as suspensory structures are of controversy,
538    though many mathematical models or drawings of the human ME include such structures,
539    cf. Table 2.1 in Mikhael, 2005; and Merchant and Nadol Jr., 2010.

540

541    It has been proposed by Rosowski et al., 1999 that the anterior mallear ligament is a bony
542    connection to the bulla, while Elkhouri et al., 2006 observed the presence of some
543    connective tissue. Our OPFOS measurements could not distinguish any soft tissue, and our
544    CT measurement showed an ossified or cartilaginous connection. The anterior process also
545    had a less pronounced shape in human than the Japanese fan-shaped structure in gerbil.

546

547    The posterior incudal ligament, which connects the incus short crus to the fossa incudes,
548    exists in many different configurations, as is illustrated in Figure 14 by Funnell, 1972 (based
549    on work by Kobayashi). From the OPFOS sections, cf. Figure 6C-E, the gerbil posterior
550    ligament appears to fall in the category of human and cat configurations. However, it is only
551    possible to appreciate the true configuration in 3D, cf. Figure 6F-G, which clearly places this
552    gerbil ligament in the category of guinea pig and rabbit. The posterior incudal ligament
553    consists of one sickle-shaped part. According to Sim and Puria, 2008, it has been observed
554    that in human the two parts shown in Figure 14 are also connected around the tip of the
555    short crus of the incus to form a single continuous ligament rather than two separate
556    ligaments. In this respect, gerbil and human ME then would be alike.

557

558    We also found the chorda tympani nerve to be present in a special arrangement in gerbil,
559    and more tightly connected to the malleus ossicle than in human: In human this nerve

560  traverses the open space of the ME cavity without actually attaching to the ossicles. In

561  gerbil, there exists a tight connection with the malleus neck and the nerve lies on top of the

562  Japanese fan-shaped anterior process sheet, cf. Figure 7 and Figure 8. Furthermore, the

563  topographic relation of the chorda tympani to the tensor tympani muscle differs from

564  human. In gerbil it runs hypotensoric (inferiorly to the tensor tympani) and in between the

565  muscle and manubrium, as was confirmed in a recent publication by Ruf et al., 2009; while in

566  human it passes epitensoric (superiorly to the tensor tympani), e.g. Maier, 2008.

567

568  We derived the ossicle bone density from our volume measurements and from mass data

569  from literature. We obtained an ossicle bone density of $1.37 \times 10^3$ kg/m³ for the stapes and

570  $1.74 \times 10^3$ kg/m³ for incus and malleus. In comparison to human, the averaged malleus

571  density is found to be $2.31 \times 10^3$ kg/m³ and the averaged incus density is $2.14 \times 10^3$ kg/m³

572  (Sim and Puria, 2008). Another source mentions an average stapes density of $2.2 \times 10^3$ kg/m³

573  in human (Kirikae, 1960; Gan et al., 2004). Hence, gerbil ossicle densities appear to be

574  significantly lower than in human.

575

576  Another contrast to human is that the stapedial artery is usually present in gerbil, while

577  seldom in human.

578

579  Finally, our observations show that the gerbil manubrium of malleus is tightly fused over its

580  full length with the TM, while in human it is mainly only fixed at the tip and lateral process of

581  the manubrium (Koike et al., 2002).

582

583

584     Resolution

585     We used state-of-the-art X-ray micro-computed tomography and the relatively new

586  orthogonal-plane fluorescence optical-sectioning microscopy on the ME. In previous CT

587  based studies of the ME, the following model resolutions were reported: 5.5 μm on gerbil

588  (Elkhouri et al., 2006),  6 μm on human (Hagr et al., 2004), 10 μm on cat (Decraemer et al.,

589  2003) and 10 μm on human (Vogel, 1999). Though these numbers are comparable to our

590  isometric 8.5 μm voxel size for μCT on gerbil bone, our data and models are of much higher

591  quality than those shown in previous work. One reason might be that the previous authors

592    stated voxel size instead of resolution, while we actually achieve a true resolution of 8.5 μm.

593    Other factors such as scan parameter settings could also account for differences in image

594    quality.

595

596    ME soft tissue imaged with medical CT devices gave poor resolution (Lemmerling et al.,

597    1997), and μCT delivered modest resolution (Sim and Puria, 2008). The same goes for MRI

598    measurements of gerbil soft tissue structures, e.g. 45 μm (Elkhouri et al., 2006). OPFOS is

599    clearly better suited to achieve high-resolution sections on ligaments and muscles – with

600    pixel sizes ranging from 1 to 5 μm – as can be seen from our sections and 3D models, cf.

601    Figure 2 and Figure 6 til Figure 11.

602

603    The OPFOS method itself is little known but it was the first setup of the growing field now

604    known as (Laser) Light Sheet based Fluorescence Microscopy (LSFM). The many different

605    implementations and improvements of the technique have been listed in a review article by

606    Buytaert et al., 2011. The construction of an OPFOS/LSFM setup is well feasible in the sense

607    that all parts needed are readily available on the market. Researchers interested in the

608    construction of such a setup or in collaboration are welcome to contact the authors, and

609    even the first commercial devices are becoming available (Buytaert et al., 2011).

610

611

612         Artifacts

613         Segmentation of the fluid-filled inner ear channels in the μCT data showed that the

614    round window in all three models is prominently bulged inwards toward the cochlea. This

615    might indicate either a small overpressure in the ME air cavity or a loss of cochlear fluid

616    because of dehydration or leakage.

617

618    Merging of OPFOS and μCT data revealed shrinking of the soft and bony tissue, most likely

619    caused by the elaborate OPFOS specimen preparation (e.g. tissue fixation, decalcification,

620    dehydration and Spalteholz treatment), though previous authors reported that this

621    procedure induced negligible shrinking (Voie, 2002; Valk et al., 2005; Hofman et al., 2008).

622    Thanks to the combination of OPFOS with μCT, we have undeformed reference data that we

623    can use to derive a scaling factor. Homogeneous scaling with 8% of the OPFOS (bone and

624    soft tissue) models has partially corrected for the shrinking artifact. After decalcification of
625    the sample, bone is reduced to a collagen matrix. The effect of decalcification cannot be
626    investigated with μCT as all calcium is removed and X-ray absorption becomes negligible. It
627    is, however, a reasonable assumption that dehydration will have a similar (and
628    homogeneous) shrinking effect on both soft tissue and *decalcified* bone. In histology, the
629    same specimen preparation (decalcification and dehydration) is performed and the same
630    assumption (homogeneous shrinkage) is adopted.

631

632    Another artifact related to specimen preparation was noticed on the stapes. The footplate of
633    the stapes is clearly convex and bulges inwards to the cochlea in the μCT models, e.g. Figure
634    10. After decalcification, dehydration and Spalteholz treatment, the footplate of the OPFOS
635    model showed some relaxation and shriveling of its convex shape. The models available for
636    download therefore consist of μCT data for bone and OPFOS data for soft tissue meshes.

637

638    *Stripe* artifacts in OPFOS were strongly reduced but not entirely eliminated by our bi-
639    directional illumination/sectioning OPFOS setup. In combination with manual segmentation,
640    which also partially corrects for this artifact, no effect remained in the models so no image
641    post-processing of the data was required.

642

643    OPFOS was not suited to image the TM, but the full-field outline of the TM shape was
644    obtained from μCT: We could not measure a volume model of the TM with the correct
645    thickness, but only a surface model. FEM modelers can, however, use the surface shape
646    directly as a shell model, cf. Gan et al., 2004; Elkhouri et al., 2006; and apply either a uniform
647    or a varying measured thickness distribution to their own choosing (as different approaches
648    are taken by different modelers). Table 2 mentions average thickness data at three TM
649    regions, measured on eleven gerbil TMs with confocal microscopy (Kuypers et al., 2005).

650

651

652        Open source availability

653        All three-dimensional data and surface mesh models presented in this paper are

654    freely available for educational and research purposes on the website of the Laboratory of

655    BioMedical Physics: http://www.ua.ac.be/bimef/models.

656

657     Several educational and research 3D models have also been made available in the past

658     (Eaton-Peabody Laboratory of Auditory Physiology, Ear & Auditory Research Laboratory,

659     Auditory Mechanics Laboratory, Auditory Research Laboratory, OtoBiomechanics Group).

660

661

662     CONCLUSION

663

664     Finite-element computer modeling needs accurate three-dimensional models to obtain

665     realistic simulation results for middle ear mechanics. 3D models are also useful in medical

666     training or for the interpretation and presentation of experimental results. The middle ear

667     does not only comprise the ossicles but also consists of soft tissue: tympanic membrane,

668     ligaments, muscles, tendon and blood vessels.

669

670     In this paper, we presented an accurate and complete morphological 3D middle (and inner)

671     ear model of gerbil. The model is freely available to the research community at our website.

672     The presented model quality is unprecedented. The position, orientation and size of all

673     components making up the gerbil middle ear are now accurately known and individually

674     discussed.

675

676

677     ACKNOWLEDGEMENTS

678

679     We gratefully acknowledge the financial support of the Research Foundation – Flanders and

680     the Fondation Belge de la Vocation. We thank Pieter Vanderniepen for his assistance in

681     operating the CT device, Magnus Von Unge and Wim Decraemer for their feedback on

682     human and gerbil anatomy, Fred Wiese for manufacturing the vial holder, Robert Funnell for

683     the use of Figure 14, and Peter Aerts for feedback on muscle functionality.

684

685

686

687

853

854

855

856

857

858

859

860

861

862

863     TABLE CAPTIONS

864

865     *Table 1: Volume, surface area and number of triangles for gerbil 1 (G1), gerbil 2 (G2) and*

866     *gerbil 3 (G3) ear components, derived from the 3D surface meshes obtained from µCT. The*

867     *ME cavity volume incorporates the air, ossicles and ME soft tissue volume.*

868

869

870     *Table 2: Geometry parameters of the TM for gerbil 1 (G1), gerbil 2 (G2) and gerbil 3 (G3) ear*

871     *components, derived from the 3D surface meshes obtained from µCT.*

872     *(*Mean thickness data from confocal microscopy on 11 gerbils by* Kuypers et al., 2005.*)*

873

874

875     *Table 3: 3D length of the manubrium (umbo tip till lateral process tip, cf. figure 5) and 3D*

876     *height of the stapes (medial footplate till tip stapes head) for gerbil 1 (G1), gerbil 2 (G2) and*

877     *gerbil 3 (G3), derived from the 3D surface meshes obtained from µCT.*

878

879

880

881

882

883

884

885

886

887

888

889

890

891

892

893

894

895    FIGURES

896

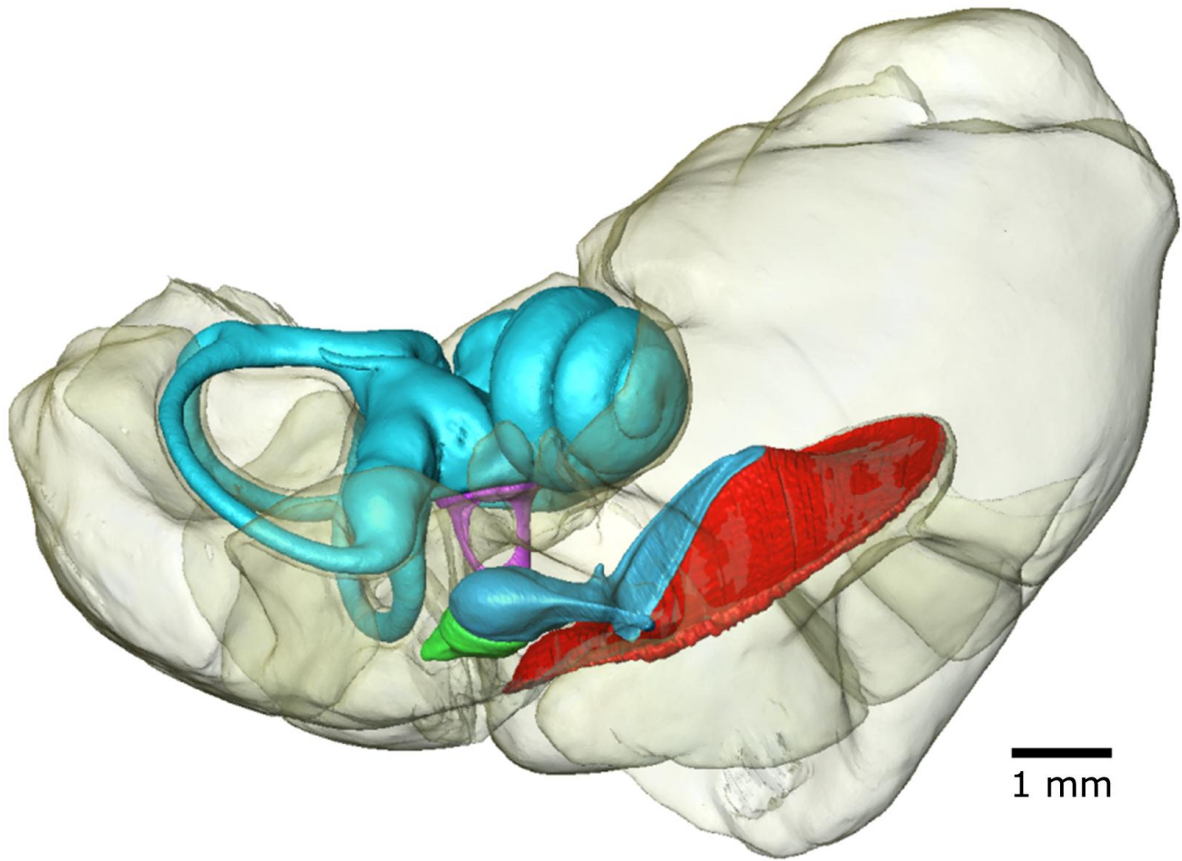

897

898    Figure 1: 3D model of separate surface meshes of bony middle and inner ear components of

899    gerbil 2, obtained from μCT. The bulla is rendered transparent. Voxel size 8.5x8.5x8.5 μm.

900

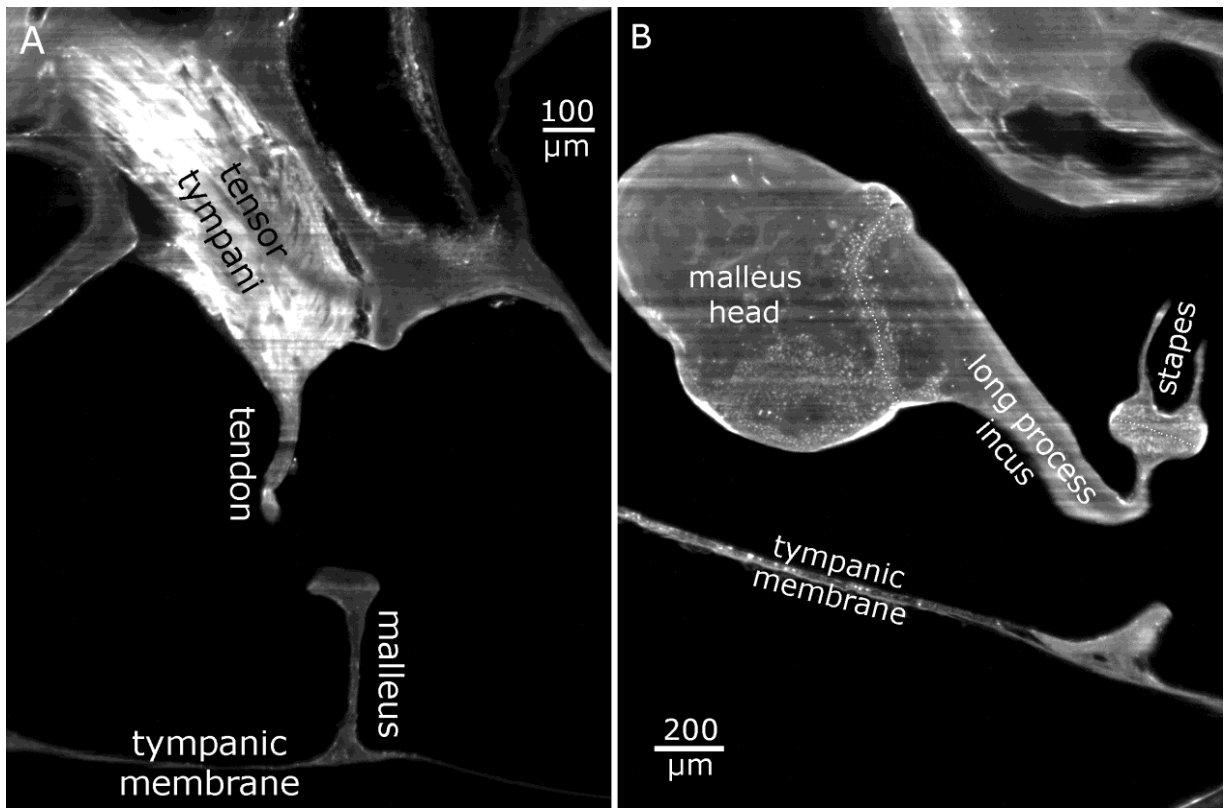

901

902 Figure 2: 2D virtual cross sections delivered by the OPFOS technique. A) Tensor tympani

903 tendon reaching down towards malleus. B) Incudomallear and incudostapedial articulation.

904 Pixel size 1.5x1.5 µm.

905

906

907

908

909

910

911

912

913

914

915

916

917

918

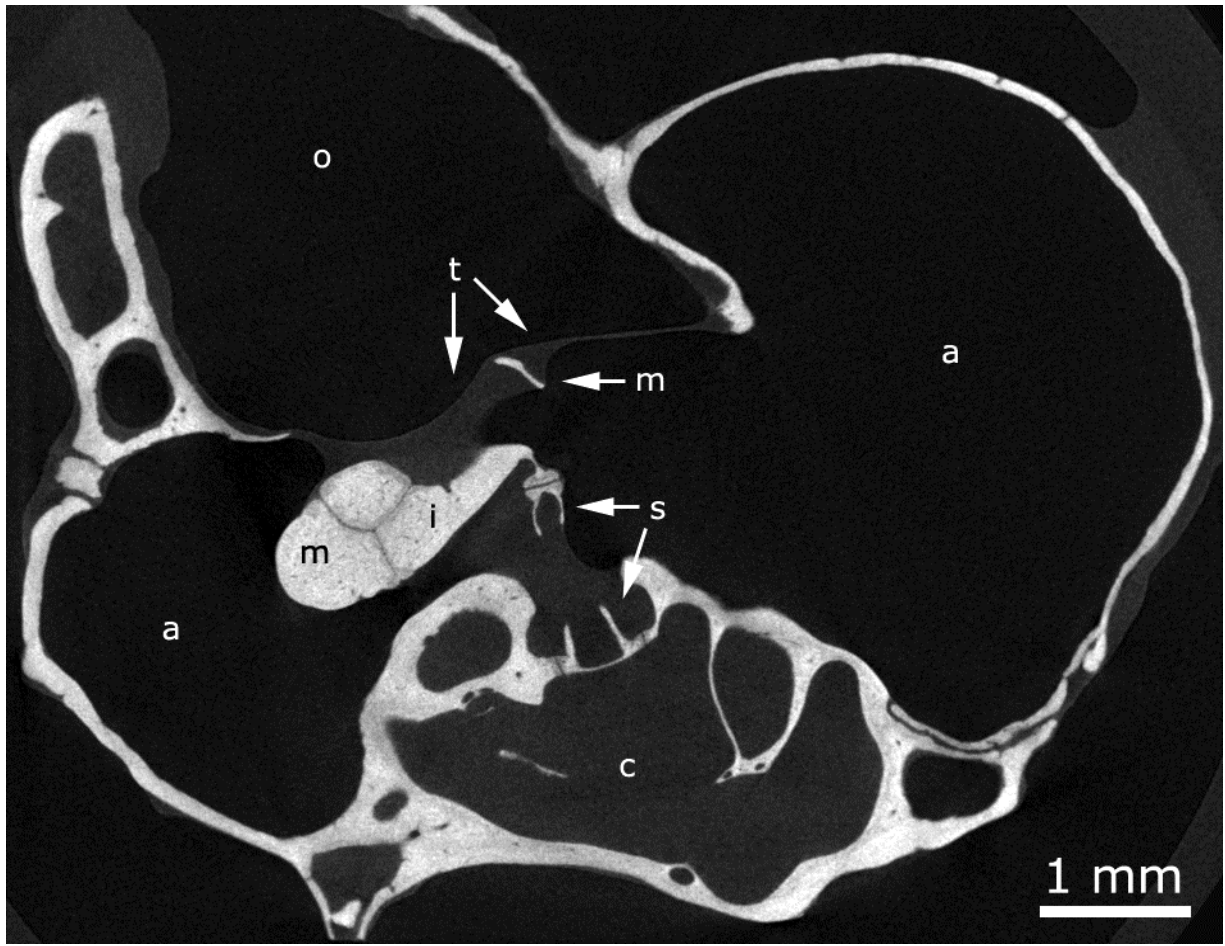

919

920    Figure 3: Reconstructed µCT cross section through gerbil 1 (originally 1496x1496 pixels

921    cropped to 740x950 pixels). a: middle ear air cavity, c: inner ear cochlea, i: incus, m: malleus,

922    o: outer ear canal, s: stapes, t: tympanic membrane outline. Pixel size 8.5x8.5 µm.

923

924

925

926

927

928

929

930

931

932

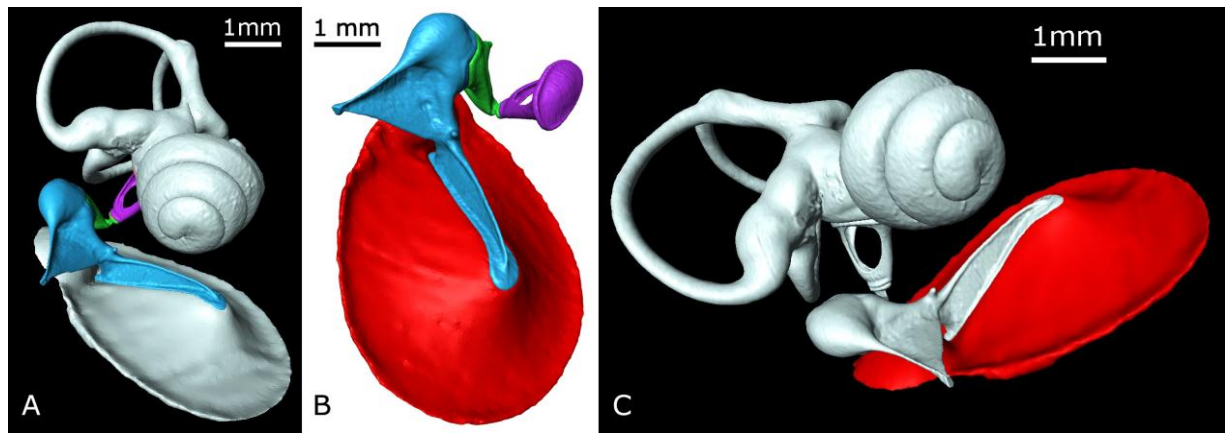

Figure 4: 3D surface meshes. Voxel size 8.5x8.5x8.5 µm.

A) Tympanic membrane + middle ear ossicles + inner ear fluid (gerbil 1).

B) Tympanic membrane + middle ear ossicles (gerbil 2).

C) Tympanic membrane + middle ear ossicles + inner ear fluid (gerbil 3).

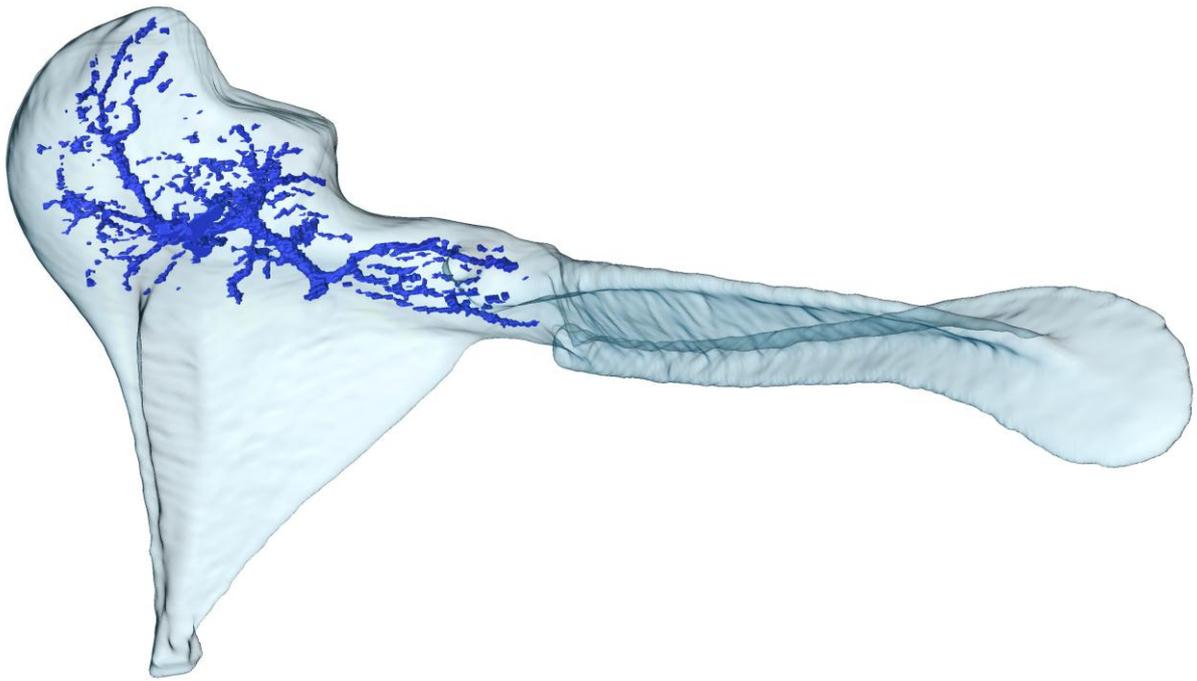

956

957  Figure 5: Mesh of the malleus (gerbil 2) rendered transparent in combination with a mesh of

958  the (major) blood vessel channels running inside it. Data obtained from µCT. Voxel size

959  8.5x8.5x8.5 µm.

960

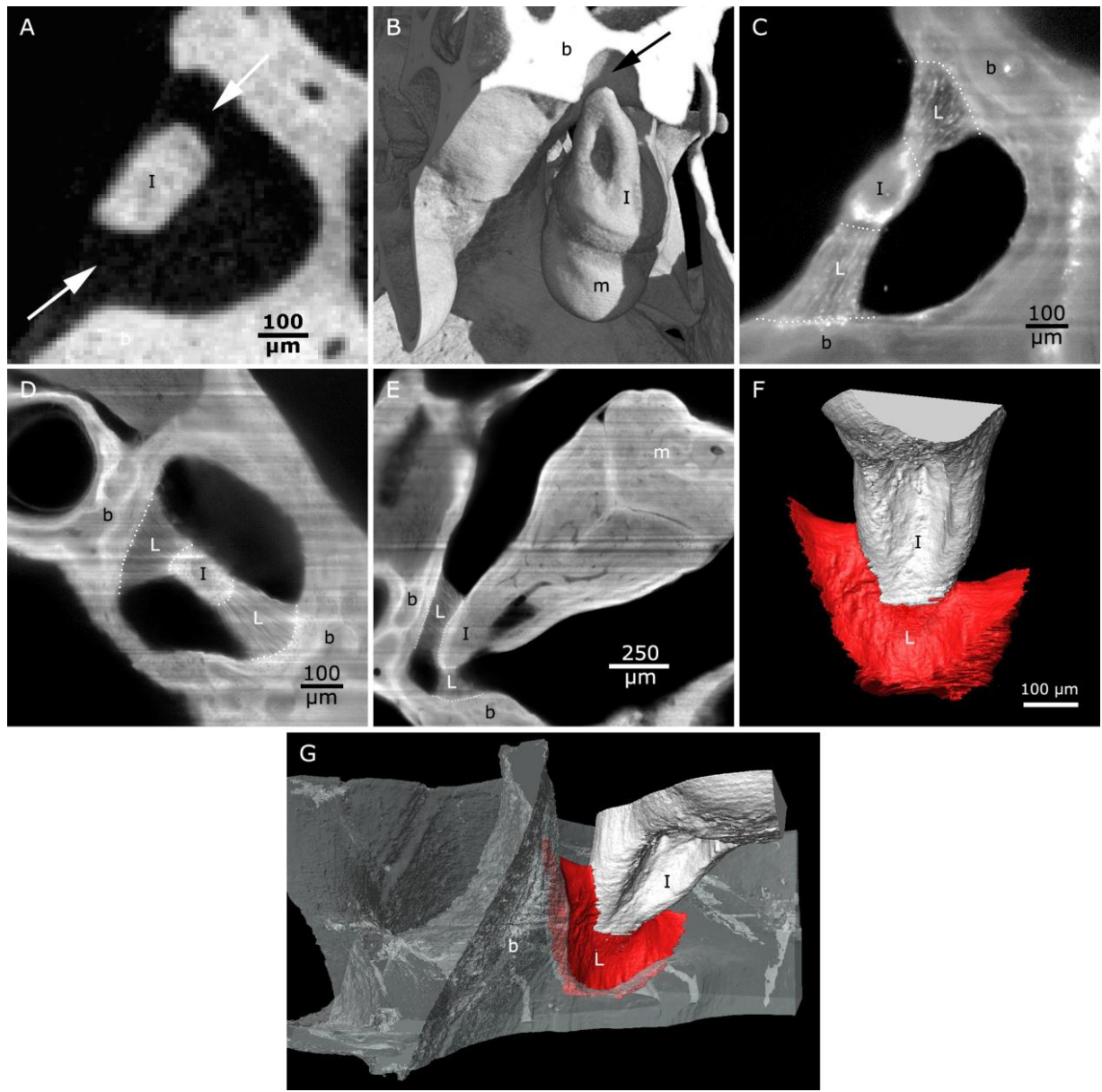

961

962    Figure 6: A) µCT cross section and B) 3D µCT reconstruction from automatic thresholding do

963    not show the posterior incudal ligament in the bony wall recess. Arrows indicate the position

964    of the invisible ligament. Pixel (and voxel) size 8.5x8.5(x8.5) µm. C-E) ROI OPFOS cross

965    sections from different orientations do show the ligament in the recess. F-G) 3D OPFOS

966    meshes. Voxel size 0.97x0.97x2.5 µm. b: bulla,  I: incus, L: ligament, m: malleus.

967

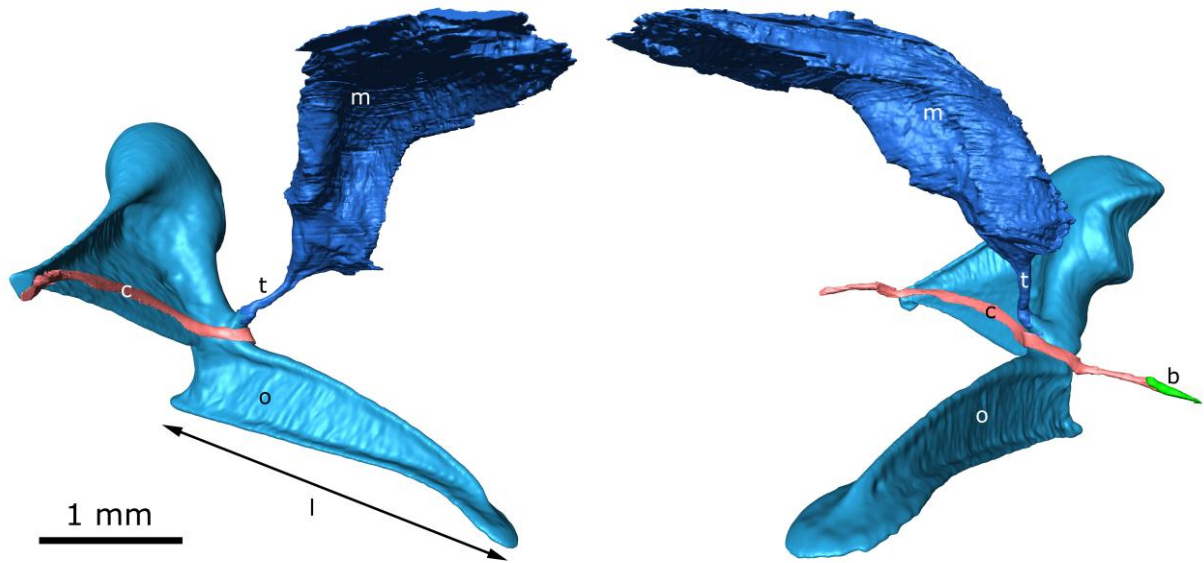

968

969 Figure 7: Two views of the topography of the chorda tympani in combination with the

970 malleus and the tensor tympani muscle & tendon (gerbil 2). The soft tissue data originate

971 from OPFOS (voxel size 2x2x4.5 µm), while the malleus data come from µCT (voxel size

972 8.5x8.5x8.5 µm). b: bulla, c: chorda tympani, m: muscle, o: malleus ossicle, t: tendon, l:

973 manubrium length.

974

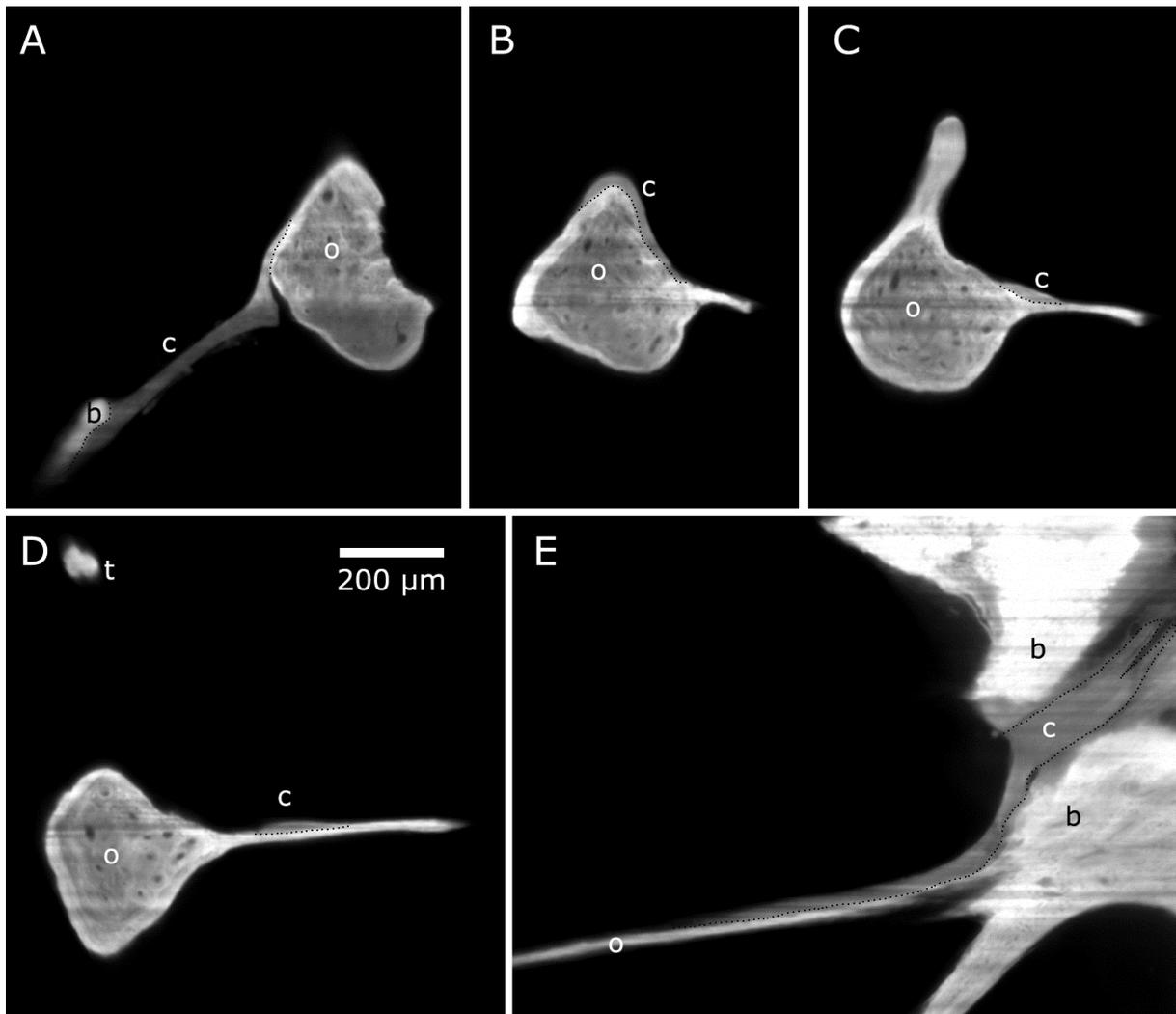

975
976 Figure 8: OPFOS cross sections showing the course of the chorda tympani with respect to the
977 malleus. A) Chorda tympani jumps from a bony support beam to the malleus neck superior
978 side. B) It rounds the malleus neck below the tensor tympani. C,D,E) It continues on the
979 anterior process sheet until it enters a fissure in the bulla wall. b: bulla, c: chorda tympani, o:
980 malleus ossicle, t: tendon. All subfigures are of the same scale.
981
982
983
984
985

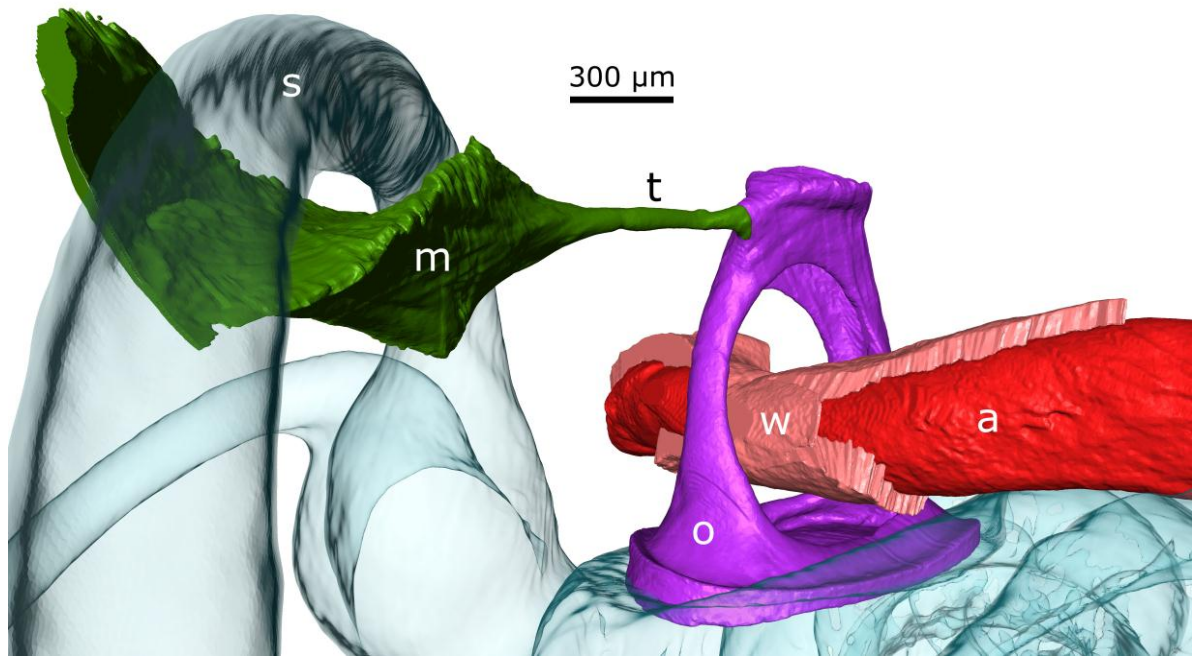

986
987    Figure 9: Stapes bone, stapedius muscle and tendon, and stapedial artery models obtained
988    from OPFOS (voxel size 1.5x1.5x5 μm), and the fluid-filled cavity of the horizontal semi-
989    circular canal from μCT (voxel size 8.5x8.5x8.5 μm) are shown (gerbil 2). a: artery,  m:
990    muscle, o: stapes ossicle, s: semi-circular canal, t: tendon, w: artery wall.
991

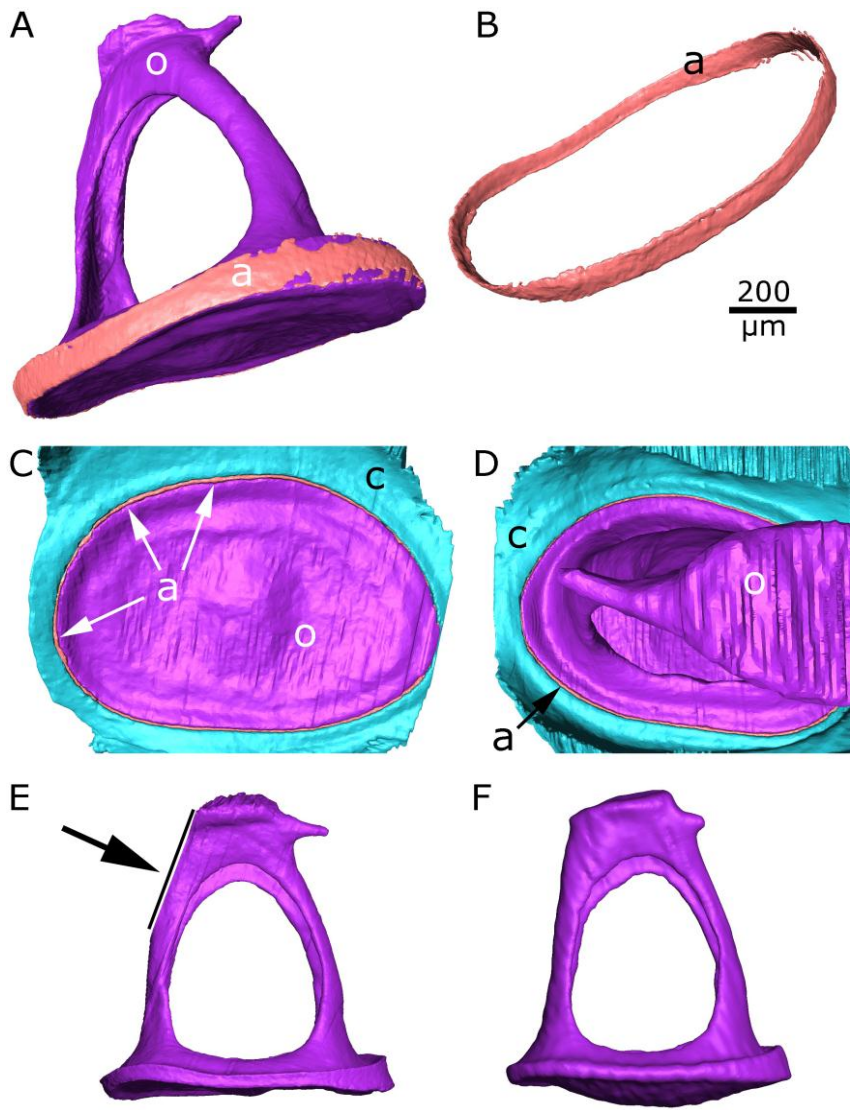

992
993   Figure 10: A-E) OPFOS based models of the stapes and the stapedial annular ligament (gerbil
994   2). F) μCT based model of the stapes (gerbil 2). The footplate modeled from μCT data is
995   convex, while in the OPFOS model it is not. a: annular ligament, c: cochlea, o: stapes ossicle.
996   The arrow indicates the end of the OPFOS dataset.
997

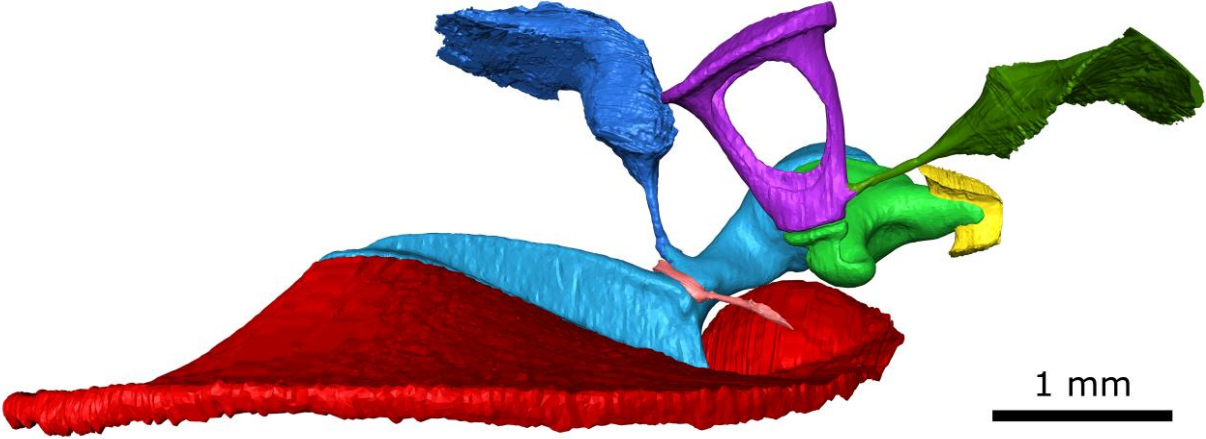

998
999  Figure 11: Merged OPFOS-CT ME model (gerbil 2).
1000
1001
1002
1003
1004
1005
1006
1007
1008
1009
1010
1011
1012
1013
1014
1015
1016
1017
1018
1019
1020
1021
1022

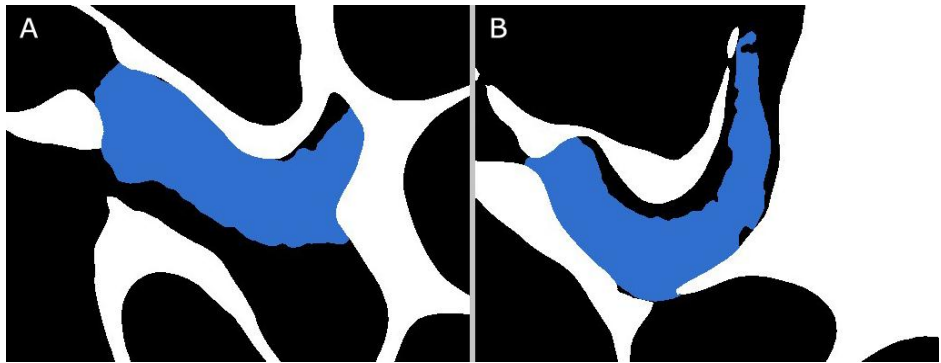

1023
1024  Figure 12: Cross sections at different depths through the 3D merged models of the bulla
1025  bone (white) from μCT and the tensor tympani (blue) from OPFOS. Black represents air filled
1026  space such as the ME air cavity. The tensor tympani fits nicely in the bone, rather touching
1027  the cavity wall than overlapping with it.
1028
1029
1030
1031
1032
1033
1034
1035
1036
1037
1038
1039
1040
1041
1042
1043
1044
1045

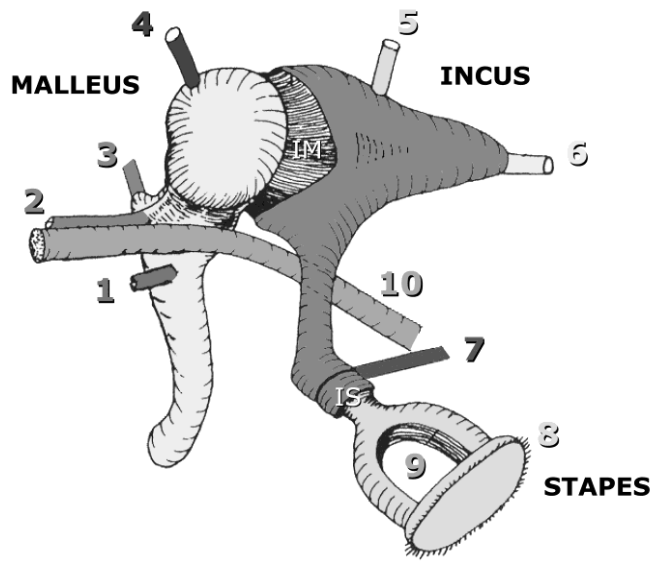

Ossicles:
  M malleus BONE
  I  incus BONE
  S  stapes BONE

Ossicle joints:
  IM and IS cleft and LIGAMENT

Ossicle suspension:
  M.1 Tensor tympani TENDON
  M.2 Anterior mallear LIGAMENT
  M.3 Lateral mallear LIGAMENT
  M.4 Superior mallear LIGAMENT
  I.5  Superior incudal LIGAMENT
  I.6  Posterior incudal LIGAMENT
  S.7 Stapedial muscle TENDON
  S.8 Stapedial annular LIGAMENT

Other soft tissue:
  S.9 Stapedial ARTERY
  M.10 Chorda tympani NERVE

1046

1047 Figure 13: General schematic overview of all relevant middle ear components in human.

1048

1049

1050

1051

1052

1053

1054

1055

1056

1057

1058

1059

1060

1061

1062

1063

1064

1065

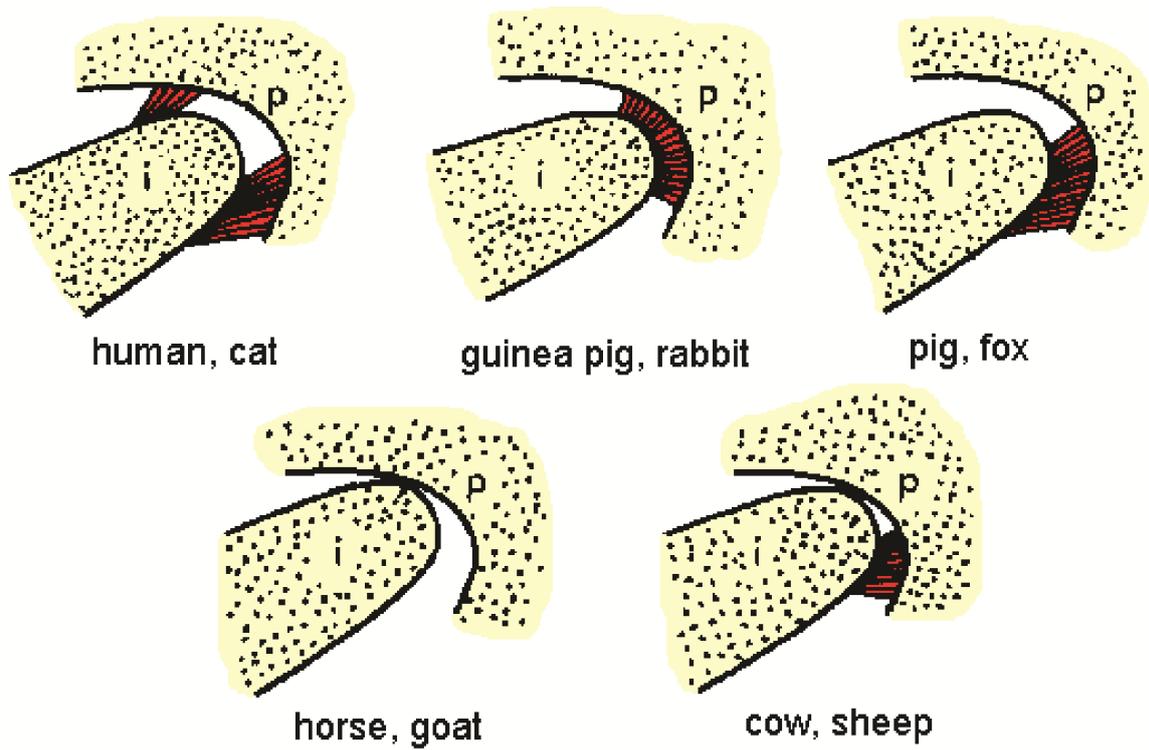

1066

Figure 14: Schematic representation of different posterior incudal ligament configurations

per species (courtesy of Funnell, 1972). Gerbil falls in the category of guinea pig and rabbit.

1069